\documentclass[aps,preprint]{revtex4}%
\usepackage{amsfonts}
\usepackage{amsmath}
\usepackage{amssymb}
\usepackage{graphicx}%
\newtheorem{theorem}{Theorem}
\newtheorem{acknowledgement}[theorem]{Acknowledgement}

\begin{document}
\title[Adsorption of Two-State Polymers]{On the Adsorption of Two-State Polymers }
\author{N. Yoshinaga}
\affiliation{Department of Physics, Graduate School of Science, the University of Tokyo,
Tokyo 113-0033, Japan.}
\author{E. Kats}
\affiliation{Laue-Langevin Institute, F-38042, Grenoble, France, and L. D. Landau Institute
for Theoretical Physics, RAS, 117940 GSP-1, Moscow, Russia}
\author{A. Halperin$^{\ast}$}
\affiliation{Structures et Propri{\'{e}}t{\'{e}}s dArchitectures Mol{\'{e}}culaires, UMR
5819 (CEA, CNRS, UJF), INAC, CEA-Grenoble, 38054 Grenoble cedex 9, France}

\begin{abstract}
Monte Carlo(MC) simulations produce evidence that annealed copolymers
incorporating two interconverting monomers, P and H, adsorb as homopolymers
with an effective adsorption energy per monomer, $\epsilon_{eff}$, that
depends on the PH equilibrium constants in the bulk and at the surface. The
cross-over exponent, $\Phi,$ is unmodified. The MC results on the overall PH
ratio, the PH ratio at the surface and in the bulk as well as the number of
adsorbed monomers are in quantitative agreement with this hypothesis and the
theoretically derived $\epsilon_{eff}$. The evidence suggests that the form of
surface potential does not affect $\Phi$ but does influence the PH equilibrium.

\end{abstract}
\volumeyear{ }
\volumenumber{ }
\issuenumber{ }
\eid{identifier}
\date{25/5/08}
\received[Received text]{date}

\revised[Revised text]{date}

\accepted[Accepted text]{date}

\published[Published text]{date}

\maketitle

\section{Introduction}

The interest in polymer adsorption reflects both its practical
importance\cite{napper} and the associated theoretical challenges\cite{ee}.
Within this domain the adsorption behavior of neutral water soluble polymers
(NWSP), exemplified by polyethylene oxide (PEO) and polyvinyl pyrrolidone
(PVP)\cite{Moly}, occupies a special position. On the applied side this is
because of their importance in the formulation of water-based colloidal
solutions\cite{Ed1, Ed2, handbook} of practical interest. From a theoretical
point of view the adsorption of NWSP poses a distinctive problem: While NWSP
such as PEO are homopolymers comprising of chemically identical monomers, they
are modeled as "two-state polymers" whose monomers interconvert between
hydrophilic (P) and hydrophobic (H) states\cite{gold, karl, tanaka, Veyt, PGG,
Bekiranov1, kremer, Bekiranov2, dorm}. These two-state models are invoked in
order to rationalize a phase diagram exhibiting a closed insolubility loop,
with both upper and lower critical solution temperatures, that is thought to
characterize NWSP. The models differ with respect to the precise
identification of the two interconverting states. Nevertheless, within all of
these models NWSP exhibit an annealed sequence of HP states with a temperature
(T) dependent H/P ratio. The adsorption behavior of such \textit{annealed}
two-state polymers is expected to differ from that of homopolymers comprising
of single monomer species and of \textit{quenched} random copolymers with a
fixed sequence and composition. Thus far, it was investigated using a self
consistent field theory \cite{Linse} of a laterally uniform adsorbed layer,
comprising many adsorbed chains, in contact with a polymer solution. In the
following we consider, in contrast, the adsorption transition of a single,
terminally anchored, two-state polymer and compare it to the corresponding
results for homopolymers \cite{ee, EKB, Metzger, Luo, Momo} and for quenched
copolymers \cite{Baum, W, W rev, Mainz, Z}. In particular, we investigate the
adsorption of a non-cooperative two-state polymer using Monte Carlo simulation
supplemented by simple theory. Our discussion is mostly concerned with swollen
chains under good solvent conditions. Simulation evidence suggests that the
adsorption of both homopolymers and of quenched copolymers exhibits identical
scaling behavior upon introduction of the appropriate effective adsorption
energy per copolymer's monomer, $\epsilon_{eff}$. The cross-over exponent
$\Phi$ characterizing the second-order adsorption transition of these two
systems is identical \cite{Baum, Z, Mainz}. With this in mind we expect
similar behavior for the annealed two-state polymers. Accordingly we first
identify the $\epsilon_{eff}$ for the annealed copolymers case and then
analyze the simulation data assuming that such chains adsorb as homopolymers
with $\epsilon_{eff}$ specifying the monomer-surface interaction. As we shall
see, this picture is consistent with the simulation results. In particular, it
allows to collapse the simulation data concerning the number of adsorbed
monomers $N_{S}$ onto universal curves and to reproduce the simulation results
concerning the total $H$ fraction as well as the $H$ fraction in the bulk and
at the surface. Importantly, in distinction to quenched copolymers, the total
$H$ fraction as well as the $H$ fraction in the bulk and at the surface are
not fixed. These results are of interest from two points of view. First, they
complement earlier results on the adsorption transition of
homopolymers\cite{ee, EKB, Metzger, Luo, Momo} and quenched
copolymers\cite{Baum, W, W rev, Mainz, Z}. Second, they provide a starting
point for the formulation of a phenomenological free energy
description\cite{Momo} of the adsorption of NWSP modeled as two-state polymers.

In formulating the problem we aimed to capture the generic features common to
the various two-state models. These differ with respect to the precise
identification of the interconverting states: polar vs apolar\cite{karl},
hydrated vs nonhydrated\cite{tanaka, Veyt, Bekiranov1, Bekiranov2, dorm}, or
clustered vs nonclustered\cite{PGG}. With this in mind we focused on the
simplest case where the monomers undergo a unimolecular HP interconversion. In
the following we confront simple theory with off-lattice Monte Carlo
simulations of the adsorption behavior of a single two-state polymer within
this minimal model. There is no explicit solvent in the simulation and the
monomers states are modeled as Lennard-Jones particles having identical
collision diameters. The interactions between of the various monomer-monomer
pairs are identical but the interaction parameters with the surface differ
with the monomeric state. As a result the PH interconvesion at the surface and
at the bulk are associated with different equilibrium constants (Figure 1).
Our model is thus closest to the one proposed by Karlstr{\"{o}}m\cite{karl}.
$\epsilon_{eff}$ of the two-state model is obtained in section II using a
partition function method similar to that of Moghaddam and Whittington\cite{W}
as well as its free energy counterpart. The consequences for the scaling
analysis and blob description are discussed in section III. In section IV we
present the details of the simulation model and the simulation results are
discussed in section V.

\section{The Effective Adsorption Energy $\epsilon_{eff}$ of Annealed
Copolymers}

Two methods can be used to identify $\epsilon_{eff},$ the adsorption energy of
an "effective" monomer at the surface. In one, first used by Moghaddam and
Whittington to obtain bounds for quenched copolymers \cite{W, W rev},
$\epsilon_{eff}$ is obtained upon recasting the annealed copolymer partition
function into a homopolymer-like form containing factors of the form
$\exp(N_{S}\frac{\epsilon_{eff}}{kT})$ where $k$ is the Boltzmann constant.
Similarly, it is possible to consider the free energy of an annealed copolymer
and recognize $\epsilon_{eff}$ in a term of the form $-N_{S}\epsilon_{eff}$.
These two equivalent methods yield as expected an identical $\epsilon_{eff}$.
We briefly discuss the two versions in order to establish the relationship
both to the Moghaddam-Whittington method and to the phenomenological free
energy approach.

We begin with the free energy of an adsorbed two-state chain. It comprises two
terms, $F_{conf}$ and $F_{\operatorname{seq}}.\ $\ The first, $F_{conf}$,
allows for the adsorption induced confinement of the chain. It reflects the
loss of configurational entropy of the flexible backbone upon confinement to a
slab of thickness $D\leqslant R_{F}$ where $R_{F}\sim N^{3/5}$ is the Flory
radius of the swollen chain$.$\ $F_{conf}$ depends only on $D,$ or
equivalently $N_{S},$ and its precise functional form is not important for the
first part of our discussion$.$ The second term, $F_{\operatorname{seq}}$,
accounts for the standard chemical potentials of the $P$ and $H$ monomers in
the bulk and at the surface as well as the mixing entropies at the surface and
in the bulk. In the good solvent regime the chains are swollen, the monomer
concentration is dilute and monomer-monomer interactions have negligible
effect on the $P\rightleftarrows H$ $\ $equilibrium. The $PH$ sequences in the
bulk and at the surface are thus modeled as ideal one-dimensional mixtures
and
\begin{equation}
F_{\operatorname{seq}}(N_{S},x_{H},x_{SH})=E-TS_{mix}=N_{B}f_{B}+N_{S}%
f_{S}=Nf_{B}+N_{S}\left(  f_{S}-f_{B}\right)  . \label{I1}%
\end{equation}
Here $f_{B}$ and $f_{S}$ are respectively the free energies per monomer in the
bulk and at the surface as specified by%

\begin{equation}
f_{S}=x_{SH}\mu_{SH}^{0}+(1-x_{SH})\mu_{SP}^{0}+kT\left[  x_{SH}\ln
x_{SH}+(1-x_{SH})\ln(1-x_{SH})\right],  \label{I2}%
\end{equation}%
\begin{equation}
f_{B}=x_{H}\mu_{H}^{0}+(1-x_{H})\mu_{P}^{0}+kT\left[  x_{H}\ln x_{H}%
+(1-x_{H})\ln(1-x_{H})\right],  \label{I3}%
\end{equation}
where the $N$ monomers in the chain comprise $N_{S}$ adsorbed monomers at the
surface and $N_{B}=N-N_{S}$ nonadsorbed monomers in the bulk. The $H$ fraction
among free monomers is $x_{H}=$ $\frac{N_{BH}}{N_{B}}$ while for the adsorbed
monomers it is $x_{SH}=\frac{N_{SH}}{N_{S}}$ where $N_{BH}$ and $N_{SH}$
denote respectively the number of $H$ monomers in the bulk and at the surface.
The standard chemical potentials of the different species are denoted by
$\mu_{i}^{0}$. Note that $f_{B}$ and $f_{S}$ as specified by (\ref{I2}) and
(\ref{I3}) are similar to the Zimm-Bragg free energy for the helix-coil
transition in the case of zero cooperativity \cite{GK}$.$ At this point
$f_{B}$ and $f_{S}$ are completely decoupled from each other and from
$F_{conf}$. The equilibrium conditions $\partial f_{B}/\partial x_{H}=0$ and
$\partial f_{S}/\partial x_{SH}=0$ lead to the mass action laws in the bulk%

\begin{equation}
\frac{N_{BH}}{N_{BP}}=\frac{x_{H}}{1-x_{H}}=K_{B}=\exp\left(  -\frac{\mu
_{H}^{0}-\mu_{P}^{0}}{kT}\right),  \label{I4}%
\end{equation}
and at the surface%

\begin{equation}
\frac{N_{SH}}{N_{SP}}=\frac{x_{SH}}{1-x_{SH}}=K_{S}=\exp\left(  -\frac
{\mu_{SH}^{0}-\mu_{SP}^{0}}{kT}\right).  \label{I5}%
\end{equation}
Accordingly, a mass action law of the form $K=\frac{x}{1-x}$ or $x=K/(1+K)$
leads, upon substitution in (\ref{I2}) and (\ref{I3}), to the equilibrium free
energies per monomer at the surface and in the bulk%

\begin{equation}
f_{S}^{eq}=\mu_{SP}^{0}-kT\ln(1+K_{S})=-kT\ln\left[  \exp\left(  -\frac
{\mu_{SP}^{0}}{kT}\right)  +\exp\left(  -\frac{\mu_{SH}^{0}}{kT}\right)
\right],   \label{I6}%
\end{equation}%
\begin{equation}
f_{B}^{eq}\equiv\epsilon_{B}=\mu_{P}^{0}-kT\ln(1+K_{B})=-kT\ln\left[
\exp\left(  -\frac{\mu_{P}^{0}}{kT}\right)  +\exp\left(  -\frac{\mu_{H}^{0}%
}{kT}\right)  \right]  . \label{I6a}%
\end{equation}
In turn, these yield $F_{\operatorname{seq}}(N_{S},x_{H},x_{SH})=Nf_{B}%
^{eq}+N_{S}\left(  f_{S}^{eq}-f_{B}^{eq}\right)  \equiv N\epsilon_{B}%
-N_{S}\epsilon_{eff}$ at equilibrium and%

\begin{equation}
\epsilon_{eff}=f_{B}^{eq}-f_{S}^{eq}=\mu_{P}^{0}-\mu_{SP}^{0}+kT\ln
\frac{(1+K_{S})}{(1+K_{B})}=kT\ln\left[  \frac{\exp\left(  -\frac{\mu_{SP}%
^{0}}{kT}\right)  +\exp\left(  -\frac{\mu_{SH}^{0}}{kT}\right)  }{\exp\left(
-\frac{\mu_{P}^{0}}{kT}\right)  +\exp\left(  -\frac{\mu_{H}^{0}}{kT}\right)
}\right].  \label{I7}%
\end{equation}
Our sign convention for $\epsilon_{eff}$ follows the custom in the
phenomenological theories of adsorption\cite{dG, Momo, RK}.

Having obtained $\epsilon_{eff}$ from the free energy of an adsorbed chain we
turn to the partition function argument. The partition function of an annealed
two-state polymer at a surface is
\[
Z=\sum_{N_{S}}\sum_{N_{SH}=0}^{N_{SH}=N_{S}}\sum_{N_{BH}=0}^{N_{BH}=N_{B}%
}C_{N}^{+}(N_{S})
\]%
\[
\times\frac{(N-N_{S})!}{N_{BH}!(N-N_{S}-N_{BH})!}\left[  \exp\left(
-\frac{\mu_{H}^{0}}{kT}\right)  \right]  ^{N_{BH}}\left[  \exp\left(
-\frac{\mu_{P}^{0}}{kT}\right)  \right]  ^{N-N_{S}-N_{BH}}%
\]%
\begin{equation}
\times\frac{N_{S}!}{N_{SH}!(N_{S}-N_{SH})!}\left[  \exp\left(  -\frac{\mu
_{SH}^{0}}{kT}\right)  \right]  ^{N_{SH}}\left[  \exp\left(  -\frac{\mu
_{SP}^{0}}{kT}\right)  \right]  ^{N_{S}-N_{SH}}. \label{II2}%
\end{equation}
Here $C_{N}^{+}(N_{S})$ is the number of chain trajectories with $N_{S}$
surface contacts which is assumed to be identical to that of a homopolymer.
$N_{SP}=N_{S}-N_{SH}$ is the number of surface monomers in $P$ state and
$N_{BP}=N-N_{S}-N_{BH}$ is the corresponding number among bulk monomers. $Z$
as given by (\ref{II2}) counts all possible free and adsorbed PH sequences and
assigns to each sequence the appropriate Boltzmann factor. The summations over
$N_{BH}$ and $N_{SH}$ yields
\[
Z=\sum_{N_{S}}C_{N}^{+}(N_{S})\left[  \exp\left(  -\frac{\mu_{H}^{0}}%
{kT}\right)  +\exp\left(  -\frac{\mu_{P}^{0}}{kT}\right)  \right]  ^{N-N_{S}}%
\]%
\begin{equation}
\times\left[  \exp\left(  -\frac{\mu_{SH}^{0}}{kT}\right)  +\exp\left(
-\frac{\mu_{SP}^{0}}{kT}\right)  \right]  ^{N_{S}}. \label{II3}%
\end{equation}
$Z_{HB}^{0}=\left[  \exp\left(  -\frac{\mu_{H}^{0}}{kT}\right)  +\exp\left(
-\frac{\mu_{P}^{0}}{kT}\right)  \right]  ^{N}=\exp\left(  -\frac{N\epsilon
_{B}}{kT}\right)  $ is the HP contribution to the bulk partition function of a
two-state chain comprising $N$ monomers. It counts the HP sequences in the
absence of a surface or in the presence of a perfectly non-adsorbing surface.
We now introduce $\widetilde{C}_{N}^{+}(N_{S})=C_{N}^{+}(N_{S})\exp\left(
-\frac{N\epsilon_{B}}{kT}\right)  $ which allows for both the backbone
configurations and the HP sequence at a non-adsorbing surface. In turn, this
yields a "homopolymer-like" $Z$%

\begin{equation}
Z=\sum_{N_{S}}\widetilde{C}_{N}^{+}(N_{S})\left[  \frac{\exp\left(  -\frac
{\mu_{SH}^{0}}{kT}\right)  +\exp\left(  -\frac{\mu_{SP}^{0}}{kT}\right)
}{\exp\left(  -\frac{\mu_{H}^{0}}{kT}\right)  +\exp\left(  -\frac{\mu_{P}^{0}%
}{kT}\right)  }\right]  ^{N_{S}}=\sum_{N_{S}}\widetilde{C}_{N}^{+}(N_{S}%
)\exp(N_{S}\frac{\epsilon_{eff}}{kT}), \label{II4}%
\end{equation}
allowing to identify $\epsilon_{eff},$ as given by equation (\ref{I7}).

The free energy formulation translates into the partition function description
via $F=-kT\ln Z$. In particular, for each set $N_{S},x_{H},x_{SH}$ the
sequence partition function is approximately $Z_{{seq}}(N_{S},x_{H}%
,x_{SH})=\exp\left[  -\frac{F_{{seq}}(N_{S},x_{H},x_{SH})}{kT}\right]  $ and
$Z_{{seq}}=\sum_{N_{S}}\sum_{x_{H}}\sum_{x_{SH}}C_{N}^{+}(N_{S})\exp\left[
-\frac{F_{{seq}}(N_{S},x_{H},x_{SH})}{kT}\right]  $ where $C_{N}^{+}(N_{S})$
containing the information on \thinspace$N_{S}$ corresponds roughly to
$\exp\left(  -\frac{F_{conf}}{kT}\right)  $. This correspondence is incomplete
because it is based on the Stirling approximation and the implicit assumption
that $1\ll N_{S}<N$, $1\ll N_{B}<N$, $1\ll N_{SH}<N_{S}$ and $1\ll
N_{BH}<N_{B}$. Alternatively, one may begin with the partition function and
obtain the free energy formulation upon replacing the factorials $x!$ in
$Z_{\operatorname{seq}}$ by $\exp(x\ln x-x).$ As we discussed, $\epsilon
_{eff}$ \ can be obtained from the minimized free energy which corresponds to
the maximal term of $Z$. Minimization of $F$ yields simultaneously
$\epsilon_{eff}$ and the equilibrium constants $K_{B}$ and $K_{S}$. $K_{B}$
and $K_{S}$ are obtainable from $Z$ upon minimizing of the individual terms
subject to the caveats noted earlier. The Moghaddam-Whittington(MW) annealed
approximation of the adsorption of quenched copolymers \cite{W} provided the
starting point of our discussion of the annealed copolymers adsorption. Within
this approximation a quenched copolymer with a fixed average $H$ fraction,
$N_{H}/N=p,$ and a quenched sequence is modeled as a copolymer with a fixed
$p$ but with an annealed sequence \cite{foot}. The two treatments differ in
two respects: (i) within the MW treatment $f_{B}=0.$ (ii) \ The standard
chemical potentials of monomers at the surface, $\mu_{SH}^{0}$ and $\mu
_{SP}^{0}$ are replaced in the MW approach by $\tilde{\mu}_{SH}^{0}=\mu
_{SH}^{0}-\ln p$ and $\tilde{\mu}_{SP}^{0}=\mu_{SP}^{0}-\ln(1-p)$. Accordingly
$F_{\operatorname{seq}}(N_{S},x_{H},x_{SH})=N_{S}f_{S}$ where $f_{S}%
=x_{SH}\tilde{\mu}_{SH}^{0}+(1-x_{SH})\tilde{\mu}_{SP}^{0}+kT[x_{SH}\ln
x_{SH}+(1-x_{SH})\ln(1-x_{SH})]$ and the equilibrium condition $\partial
f_{S}/\partial x_{SH}=0$ leads to the mass action law $\frac{x_{SH}}{1-x_{SH}%
}=\widetilde{K}_{S}=\frac{p}{1-p}K_{S}$ and to equilibrium $x_{SH}%
=\frac{pK_{S}}{1-p+pK_{S}}$ instead of the $x_{SH}=\frac{K_{S}}{1+K_{S}}$ as
obtained for the fully annealed polymer considered by us. For the particular
case considered by MW, where $\mu_{SP}^{0}=0$ and ${\mu}_{SH}^{0}%
/kT=-\epsilon_{MW},$ this leads to%
\begin{equation}
\frac{\epsilon_{eff}}{kT}=-\frac{f_{S}^{eq}}{kT}=\ln\left[  \exp\left(
-\frac{\widetilde{\mu}_{SP}^{0}}{kT}\right)  +\exp\left(  -\frac
{\widetilde{\mu}_{SH}^{0}}{kT}\right)  \right]  \text{ }=\ln\left[
p\exp\epsilon_{MW}+1-p\right],  \label{WM}%
\end{equation}
while the partition function (\ref{II2}) is replaced by its MW counterpart%
\begin{equation}
Z=\sum_{N_{S}}\sum_{N_{SH}=0}^{N_{SH}=N_{S}}C_{N}^{+}(N_{S})\frac{N_{S}%
!}{N_{SH}!(N_{S}-N_{SH})!}p^{N_{SH}}(1-p)^{N_{S}-N_{SH}}\exp\left(
N_{SH}\epsilon_{MW}\right).  \label{WM1}%
\end{equation}
One may thus consider the MW partition function as a special case of equation
(\ref{II3}) in which ${\mu}_{SH}^{0}/kT=-\epsilon_{MW}-\ln p$, ${\mu}_{SP}%
^{0}/kT={\mu}_{P}^{0}/kT=-\ln(1-p)$ and ${\mu}_{H}^{0}/kT=-\ln p$. The two
procedures also differ because the MW approach is an approximation when
applied to quenched copolymers while its counterpart, as described above, is
rigorously applicable to the annealed the two-state polymers considered here
\cite{foot2}.

\section{ Scaling Analysis and Adsorption Blobs}

Having identified $\epsilon_{eff}$ of two-state polymers we are in a position
to discuss the scaling analysis of such polymers and the corresponding blob
picture. For two-state polymers it assumes, as is the case for quenched
copolymers, that the polymers adsorb as homopolymers but with an excess
adsorption energy per monomer of%

\begin{equation}
\epsilon kT=\epsilon_{eff}-\epsilon_{eff}^{c}. \label{III1}%
\end{equation}
Here $\epsilon_{eff}^{c}$ is a constant, model dependent, critical adsorption
energy at the limit of $\tau=\left(  \epsilon_{eff}-\epsilon_{eff}^{c}\right)
/\epsilon_{eff}^{c}\rightarrow0,$ $N\rightarrow\infty$ while $\epsilon
N^{\Phi}=const^{\prime}$\cite{EKB}. In contrast to simple homopolymers where
$\epsilon\approx const^{\prime}$ \cite{Momo} the $T$ dependence of two-state
polymers $\epsilon$ is
\begin{equation}
\epsilon(T)=\ln\frac{\exp\left(  -\frac{\mu_{SH}^{0}}{kT}\right)  +\exp\left(
-\frac{\mu_{SP}^{0}}{kT}\right)  }{\exp\left(  -\frac{\mu_{H}^{0}}{kT}\right)
+\exp\left(  -\frac{\mu_{P}^{0}}{kT}\right)  }-\frac{\epsilon_{eff}^{c}}{kT}.
\label{III2}%
\end{equation}
Upon identifying $\epsilon,$ the "homopolymer-like" scaling hypothesis for
$N_{S}$ is standard. In particular%

\begin{equation}
N_{S}\approx N^{\Phi}g_{s}(x)\text{ where }x=\tau N^{\Phi}, \label{III3}%
\end{equation}
where $g_{s}(x)$ is a universal scaling function\cite{EKB, Metzger, Momo}. At
the transition, where $N_{S}\approx N^{\Phi},$ we thus require $g_{s}%
(x=0)\approx1$. In the adsorption region, $x>>1,$ $g_{s}(x)$ follows a power
law behavior, $x^{q_{s}}.$ Since $N_{S}\sim N$ when $x>>1$ this leads to
$q_{s}=(1-\Phi)/\Phi$ or $N_{S}\approx N\epsilon^{(1-\Phi)/\Phi}.$ Overall, a
plot of $N_{S}/N^{\Phi}$ vs $\tau N^{\Phi}$ should collapse the data onto a
single curve corresponding to $g_{s}(\tau N^{\Phi})$. To this end it is first
necessary to determine the model dependent $\epsilon_{eff}^{c}.$ For
$\Phi=1/2,$ the currently accepted value,
\begin{equation}
\frac{N_{S}^{2}}{N}\sim N^{2\Phi-1}\sim N^{0}, \label{III3a}%
\end{equation}
at the critical point, $\epsilon_{eff}=\epsilon_{eff}^{c}.$ This indicates
that curves of $N_{S}^{2}/N$ vs. $\epsilon_{eff}$ for different $N$ values
intersect at $\epsilon_{eff}=\epsilon_{eff}^{c}$ thus allowing to determine
$\epsilon_{eff}^{c}$ from the intersection point\cite{Hsu}.

We now return to the free energy per adsorbed chain, $F$. It comprises of two
terms $F=F_{conf}+F_{\operatorname{seq}}.$ The adsorption free energy
$F_{\operatorname{seq}}\approx-N_{S}\epsilon kT$ allows for the attractive
monomer-surface contacts. $F_{conf}$ reflects the confinement of the polymer
into a slab of thickness $D<R_{F}.$ Within the blob picture\cite{Momo}
$F_{conf}\approx kTN_{blob}\approx kT\frac{N}{g}$where $N_{blob}$ is the
number of confinement "$D$ blobs" comprising each of $g$ monomers such that
$g^{\nu}a\approx D$. Since $N_{S}\approx\frac{N}{g}g^{\Phi}$ we obtain
$g\approx\left(  \frac{N_{S}}{N}\right)  ^{\frac{1}{\Phi-1}}$ and thus%
\begin{equation}
\frac{F(N_{S})}{kT}\approx N\left(  \frac{N_{S}}{N}\right)  ^{\frac{1}{1-\Phi
}}-N_{S}\epsilon, \label{III4}%
\end{equation}
minimization with respect to $N_{S}$ than yields the equilibrium $N_{S}$
\begin{equation}
N_{S}\approx N\epsilon^{\frac{1-\Phi}{\Phi}}.\label{III5}%
\end{equation}
and the equilibrium free energy of the adsorbed chain%
\begin{equation}
\frac{F_{eq}}{kT}\approx-N\epsilon^{\frac{1}{\Phi}}.\label{III5b}%
\end{equation}
In turn, the equilibrium $N_{S}$ together with $x_{H}=K_{B}/(1+K_{B})$ and
$x_{SH}=K_{S}/(1+K_{S})$ determine the\ total number of $H$ monomers
$N_{H}=N_{S}\frac{K_{S}}{1+K_{S}}+\left(  N-N_{S}\right)  \frac{K_{B}}%
{1+K_{B}}$ or
\begin{equation}
\frac{N_{H}}{N}\approx\epsilon^{\frac{1-\Phi}{\Phi}}\frac{K_{S}}{1+K_{S}%
}+\left(  1-\epsilon^{\frac{1-\Phi}{\Phi}}\right)  \frac{K_{B}}{1+K_{B}%
}. \label{III5a}%
\end{equation}
\qquad

As noted earlier, the recent consensus regarding the cross-over exponent
suggests $\Phi=1/2$ \cite{Metzger}. The corresponding adsorption blobs for
three dimensional, self-avoiding chains are specified by%
\begin{equation}
g\approx1/\epsilon^{2}\text{ \ ; }D\approx a/\epsilon^{6/5}\ \text{; }%
N_{S}\approx N\epsilon. \label{III6}%
\end{equation}
However, simulation results suggest that finite size effects are important and
for finite chains $\Phi\approx3/5$ is more realistic \cite{Metzger}. For
relatively short chains the appropriate blob picture may thus involve%

\begin{equation}
g\approx1/\epsilon^{5/3}\text{ ; \ }D\approx a/\epsilon\text{ ;}N_{S}\approx
N\epsilon^{2/3}. \label{III8}%
\end{equation}

For the adsorption of free chains, the form of $\epsilon_{eff}$ leads to
distinctive adsorption constant in dilute surface regime, when the adsorbed
chains do not overlap. In this regime the chemical potential of the adsorbed
chains is $\mu_{ads}\approx F_{eq}+kT\ln\Gamma$ where $\Gamma$ is the activity
of the adsorbed chains and $F_{eq}\approx-kTN\epsilon^{\frac{1}{\Phi}}$ is the
standard chemical potential of an adsorbed chain. The chemical potential of
the free chains in the bulk is $\mu_{bulk}\approx kT\ln c_{bulk}$ where
$c_{bulk}$ is the activity of the free chains. The adsorption isotherm, as
obtained from $\mu_{ads}=\mu_{bulk},$ is $\Gamma\approx c_{bulk}K_{ads}$ where
$K_{ads}\approx\exp(N\epsilon^{\frac{1}{\Phi}})$ is the adsorption
constant\cite{Momo}. $K_{ads}$ for two state polymers and $\Phi=1/2$ assumes
thus the form
\begin{equation}
K_{ads}=\exp N\left\{  \ln\frac{\exp\left(  -\frac{\mu_{SH}^{0}}{kT}\right)
+\exp\left(  -\frac{\mu_{SP}^{0}}{kT}\right)  }{\left[  \exp\left(  -\frac
{\mu_{H}^{0}}{kT}\right)  +\exp\left(  -\frac{\mu_{P}^{0}}{kT}\right)
\right]  \exp\left(  \frac{\epsilon_{eff}^{c}}{kT}\right)  }\right\}  ^{2}.
\label{III9}%
\end{equation}
The $T$ dependence of $K_{ads}$ as given by (\ref{III9}) differs from that of
$K_{ads}=\exp\left(  N\delta^{1/\Phi}\right)  $ as obtained for homopolymer
adsorption with $\epsilon=\delta\approx const^{\prime}$\cite{Momo}.

As we shall see in section IV the approach discussed above accounts well for
the simulation results. Confrontation with experimental results is more
difficult. The four parameters involved, $\mu_{H}^{0}$, $\mu_{P}^{0}$,
$\mu_{SP}^{0}$ and $\mu_{SH}^{0}$ are specific both to the particular NWSP
considered and to the model used in order to analyze the data. While $\mu
_{H}^{0}$, $\mu_{P}^{0}$ were determined for a number of two-state models, the
full set of parameters was only determined for PEO within the Karlstr{\"{o}}m
model by fitting phase boundaries and adsorption data\cite{karl,
Linse}.\ However, while this model is closest to the one we analyze, the two
differ in a number of points. For example, in contrast to our model the PP, HH
and PH interactions within the Karlstr{\"{o}}m model are not identical. With
this caveat in mind, this set of $\mu_{H}^{0}$, $\mu_{P}^{0}$, $\mu_{SP}^{0}$
and $\mu_{SH}^{0}$ can be used to illustrate the behavior of $\epsilon_{eff}$
for a "PEO-like" case$.$ In the Karlstr{\"{o}}m model the standard chemical
potential of the two-states in the bulk is given by $\mu_{i}^{0}=U_{i}-RT\ln
g_{i}$ where $U_{i}$ is the internal energy and $g_{i}$ is a degeneracy
factor. The values per mole in KJ are $\mu_{P}^{0}=0$ and $\mu_{H}%
^{0}=5.086-RT\ln8$\cite{karl}$.$ For the adsorbed species $\mu
_{\operatorname{Si}}^{0}=\mu_{i}^{0}-\Delta\epsilon_{i}^{ad}$ where the
adsorption energy per mole $\Delta\epsilon_{i}^{ad}$ at methylated silica are
$\Delta\epsilon_{P}^{ad}=0.825KJ$ and $\Delta\epsilon_{H}^{ad}=1.625KJ$
\cite{Linse}. A rough idea concerning the physical consequences of the
Karlstr{\"{o}}m model may be gained upon comparing the $\epsilon_{eff}$ of a
hypothetical P homopolymer, corresponding to PEO modeled as hydrophilic
one-state polymer, to the annealed PH homopolymer model of PEO. Such
comparison shows that $\epsilon_{eff}$ of the two-state chain is shifted
upwards by roughly a factor of two.

\section{Simulation Model}

We simulate a two-state polymer terminally anchored to a planar surface.
Following Baumgartner \cite{Baum1} the polymer is modeled as freely jointed
chain comprising $N$ Lennard-Jones (LJ) particles. The monomers within this
bead-spring model interact via a LJ potential
\begin{equation}
V_{\mathrm{LJ}}=4\widetilde{\epsilon}\sum_{i,j}\left[  \left(  \frac{a}%
{r_{ij}}\right)  ^{12}-\left(  \frac{a}{r_{ij}}\right)  ^{6}\right]
,\label{IV1}%
\end{equation}
where $r_{ij}=|\mathbf{r_{i}}-\mathbf{r_{j}}|$ is the distance between
monomers $i$ and $j$ such that $|i-j|\geq2$ and $\mathbf{r}_{i}$ is the
position vector of the $i$th monomer. $\widetilde{\epsilon}$ specifies the
depth of the potential minimum at $r=2^{1/6}a$ and $a$, the collision
diameter, is the separation for which $V_{\mathrm{LJ}}=0.$ This potential
exhibits a soft-core steric repulsion at $r\leq a$, and steeply decaying
attraction for $r\geq a$. The monomers exist in two interconverting states,
$P$ and $H$. It is thus necessary to specify three LJ potentials,
corresponding to the interactions between $PP$, $HH$, and $PH$ monomer pairs,
involving altogether six parameters. In the following all three potentials are
characterized by the same $a$ and we will thus express all distances in these
units. The three remaining parameters, $\widetilde{\epsilon}=\widetilde
{\epsilon}_{\mathrm{PP}}kT$; $\widetilde{\epsilon}_{\mathrm{PH}}kT$;
$\widetilde{\epsilon}_{\mathrm{HH}}kT$ determine the strength of interactions
between the $PP$, $PH$, and $HH$ monomers. For the simulation of self avoiding
chains we set $\widetilde{\epsilon}_{\mathrm{PH}}=\widetilde{\epsilon
}_{\mathrm{HH}}=\widetilde{\epsilon}_{\mathrm{PP}}=0.20$ so that the
monomer-monomer interactions are dominated by the excluded volume and contain
effectively no attractive contribution. For simulations of ideal polymers
$\widetilde{\epsilon}_{\mathrm{PH}}=\widetilde{\epsilon}_{\mathrm{HH}%
}=\widetilde{\epsilon}_{\mathrm{PP}}=0.$~In addition, the bulk $HP$ states are
characterized by standard chemical potentials $\mu_{H}^{0}=kT\Delta\epsilon$
and $\mu_{P}^{0}=0$ with $\Delta\epsilon>0$. Accordingly $\Delta\mu^{0}%
=\Delta\epsilon kT$ specifies the difference in standard chemical potential
$\Delta\mu^{0}=\mu_{H}^{0}-\mu_{P}^{0}$ between noninteracting $H$ and $P$
monomers. All monomer pairs, except nearest-neighbors, interact via LJ
potentials. Nearest-neighbor monomers along the chain are constrained to
$0.7\leq(\mathbf{r}_{i+1}-\mathbf{r}_{i})^{2}\leq2.0$. Separations outside
this range incur an infinite energy penalty thus ensuring connectivity.

We have used two monomer-surface interaction potentials, $U_{\mathrm{wall}}$,
both specified in terms of $z$, the distance between the monomer and the
surface. One is the contact potential
\begin{equation}
U_{\mathrm{wall}}=%
\begin{cases}
\infty & z\leq0\\
-kT\Delta\epsilon_{i}^{ad} & 0<z\leq1\\
0 & z>1
\end{cases}
\label{IV2}%
\end{equation}
where $\Delta\epsilon_{i}^{ad}>0$ is the depth of the square well adjacent to
the surface as experienced by species $i=P,H$. We will mostly focus on the
contact potential because it allows for unambiguous definition of adsorption
i.e., a monomer having its center within the slab $0<z\leq1.0$ is adsorbed.
The second, more physical, "10-4" wall potential
\begin{equation}
U_{\mathrm{wall}}=kT\Delta\epsilon_{i}^{ad}\left[  \left(  \frac{1}{z}\right)
^{10}-2.5\left(  \frac{1}{z}\right)  ^{4}\right]  , \label{IV3}%
\end{equation}
as obtained by integrating the LJ potentials between the top monolayer atoms
of the substrate and a monomer at $z$. Here again $\Delta\epsilon_{i}^{ad}>0$
is a tuning parameter characterizing the strength of the
attraction\cite{Metzger}.\ When using the 10-4 potential the criterion for
adsorption is somewhat arbitrary and we define monomers with $z<2$ as
adsorbed\cite{Metzger}. Note however that monomers within the adsorption slab
experience a $z$ dependent $U_{\mathrm{wall}}.$ In both cases the values of
$\Delta\epsilon_{i}^{ad}$ for the P and H states, $\Delta\epsilon_{H}^{ad}$ or
$\Delta\epsilon_{P}^{ad},$ differ favoring the surface H state \textit{i.e.,}
$\Delta\epsilon_{P}^{ad}<\Delta\epsilon_{H}^{ad}.$ For the contact surface
potentials we thus have%

\begin{equation}
\mu_{SH}^{0}/kT=\mu_{H}^{0}-\Delta\epsilon_{H}^{ad}=\Delta\epsilon
-\Delta\epsilon_{H}^{ad}\text{ and}~\mu_{SP}^{0}/kT=\mu_{P}^{0}-\Delta
\epsilon_{P}^{ad}=-\Delta\epsilon_{P}^{ad}. \label{IV4}%
\end{equation}
For the 10-4 potentials $\mu_{SH}^{0}$ and $~\mu_{SP}^{0}$ are $z$ dependent.
However, due to the fast decay of the potential $\mu_{SH}^{0}$ and $~\mu
_{SP}^{0}$ as given by (\ref{IV4}) provide, as we shall see, a reasonable approximation.

The simulations involves chains comprising $N=64,128,256$ monomers. At each
Monte Carlo step (MCs) we shift the position of every monomer in the chain and
update its HP state using the Metropolis algorithm. This procedure is thus
repeated $N$ times per MCs. For each set of parameters the simulation involves
$2\times10^{7}$ MCs. The system is equilibrated during million MCs. The
remaining $1.9\times10^{7}$ MCs are grouped into sets of $10^{4}$ MCs whose
configurational and PH characteristics are averaged for analysis. In a
conformational update of the monomer position, a monomer is chosen randomly,
its position is shifted by a sufficiently small distance and the resulting
energy difference $kT\Delta E$, accounting for LJ interactions, is calculated.
When $\Delta E\leq0$, the operation is accepted and we proceed to the next
monomer movement. On the other hand when $\Delta E>0$, the operation is
accepted with the probability $\exp(-\Delta E)$. The PH\ interconversion
updates are implemented in two stages corresponding to the bulk equilibration
and its modification by the surface potential. In the "bulk stage" a monomer
is randomly chosen and its state is updated with the probability
\begin{equation}%
\begin{cases}
p(\mathrm{P\rightarrow H}) & =\exp(-\Delta\epsilon)\\
p(\mathrm{H\rightarrow P}) & =1
\end{cases}
\label{IV5}%
\end{equation}
where $p(\mathrm{P\rightarrow H})$ and $p(\mathrm{H\rightarrow P})$ denote
respectively the transition probabilities from $\mathrm{P}$ to $\mathrm{H}$,
and from $\mathrm{H}$ to $\mathrm{P}$. The detailed balance condition for the
bulk equilibrium is $p(\mathrm{P\rightarrow H})N_{BP}=p(\mathrm{H\rightarrow
P})N_{BH}$ \ yields
\begin{equation}
\frac{N_{BH}}{N_{BP}}=\frac{x_{H}}{1-x_{H}}=\exp(-\Delta\epsilon)\equiv K_{B}.
\label{IV6}%
\end{equation}
This procedure is the counterpart of the trial motion in the conformational
steps. The PH states are subsequently updated, in $z$ dependent fashion,
allowing for the effect of the surface potential on the PH equilibrium. For
the contact potential, monomers with $z=1$ are converted following the
transition probabilities%

\begin{equation}%
\begin{cases}
P(S\mathrm{P\rightarrow SH}) & =\exp(\Delta\epsilon_{P}^{ad}-\Delta
\epsilon_{H}^{ad})\\
P(S\mathrm{H\rightarrow SP}) & =1
\end{cases}
\label{IV7}%
\end{equation}
where $P(S\mathrm{H\rightarrow SP})=1$ because $\Delta\epsilon_{P}^{ad}%
<\Delta\epsilon_{H}^{ad}$. The corresponding detailed balance condition for
equilibrium at the surface, $p(\mathrm{P\rightarrow H})P(SP\rightarrow
SH)N_{SP}=p(\mathrm{H\rightarrow P})P(SH\rightarrow SP)N_{SH}$ yields%

\begin{equation}
\frac{N_{SH}}{N_{SP}}=\frac{x_{SH}}{1-x_{SH}}=\exp(\Delta\epsilon_{H}%
^{ad}-\Delta\epsilon-\Delta\epsilon_{P}^{ad})=K_{S}. \label{IV8}%
\end{equation}
In contrast, monomers at $z>1$ experience no surface effect and $K_{B}$ is
accordingly unmodified. Similar procedure is used for the 10-4 wall potential
with the distinction that the $z$ dependent surface transition probabilities
occur, in principle at all $z$ values with subsequent effect on $K_{B}$ as
well as on $K_{S}$. However, since the 10-4 potential decays fast the effect
on the bulk equilibrium is negligible.

\section{Simulation Results}

As opposed to the case of quenched copolymers, whose composition is fixed, the
$H$ fraction of our model two-state polymers depends on three tuning
parameters $\Delta\epsilon,\Delta\epsilon_{H}^{ad}$ and $\Delta\epsilon
_{P}^{ad}$. In the following $\Delta\epsilon_{P}^{ad}$ is fixed while
$\Delta\epsilon$ and $\Delta\epsilon_{H}^{ad}$ are varied. Our results are
mostly concerned with chains exhibiting self avoiding random walk (SAW)
statistics and interacting with the surface via contact potentials. In
addition we will comment briefly on the behavior of ideal chains exhibiting
random walk (RW) statistics and the effect of 10-4 monomer-surface potentials.
Importantly, variation of $\Delta\epsilon_{H}^{ad}$ affects $N_{SH\text{ }}$
(Figure 2), $N_{H\text{ }}$ (Figure 3) and $N_{S}$ (Figure 4).\ As seen from
Fig 2, $N_{SH}/N_{S}=K_{S}/(1+K_{S})$ and $N_{BH}/N_{B}=K_{B}/(1+K_{B})$ in
agreement with equations (\ref{I4})\ and (\ref{I5}). To proceed further it is
first necessary to determine $\epsilon_{eff}^{c}$. As noted earlier, curves of
$N_{S}^{2}/N$ vs. $\epsilon_{eff}$ for different $N$ values intersect at
$\epsilon_{eff}=\epsilon_{eff}^{c}$ provided that $\Phi=1/2.$ Alternatively,
curves of $N_{S}^{2}/N$ vs. $\Delta\epsilon_{H}^{ad}$ will intersect at
$\Delta\epsilon_{H}^{c}$ for families of curves of identical $\Delta
\epsilon,\Delta\epsilon_{P}^{ad},$ but different $N$ values. Within the
accuracy of our data the curves do intersect at a single point (Figure 5) thus
lending support to the assumption that the cross-over exponent $\Phi=1/2$
describes the adsorption behavior of two-state polymers. Furthermore,
$\epsilon_{eff}^{c}$ as obtained by substituting $\Delta\epsilon_{H}^{c}$ and
the corresponding $\Delta\epsilon,\Delta\epsilon_{P}^{ad}$ values into
(\ref{I7}) is a constant (Figure 6) as expected within the picture of
"homopolymer-like" adsorption. Having obtained $\epsilon_{eff}^{c}$ we find
that the $N_{S}$ vs $\Delta\epsilon_{H}^{ad}$ data collapses onto a universal
curve (Figure 7) upon plotting $N_{S}/N^{1/2}$ vs $\tau N^{1/2}$ where%

\begin{equation}
\tau=\frac{kT}{\epsilon_{eff}^{c}}\ln\frac{\exp\left(  \Delta\epsilon_{H}%
^{ad}-\Delta\epsilon\right)  +\exp\left(  \Delta\epsilon_{P}^{ad}\right)
}{1+\exp\left(  -\Delta\epsilon\right)  }-1. \label{V1}%
\end{equation}
Furthermore, the $N_{H}/N$ data agrees with equation (\ref{III5a}) as adapted
to the simulation model (Figure 3)%

\begin{equation}
\frac{N_{H}}{N} \approx\epsilon\frac{\exp(\Delta\epsilon+\Delta\epsilon
_{P}^{ad}-\Delta\epsilon_{H}^{ad})}{\exp(\Delta\epsilon+\Delta\epsilon
_{P}^{ad}-\Delta\epsilon_{H}^{ad})+1}+\left(  1-\epsilon\right)  \frac
{\exp(\Delta\epsilon)}{\exp(\Delta\epsilon)+1}. \label{V2}%
\end{equation}
The above results concern SAW and contact potentials. The $\epsilon_{eff}^{c}$
values for RW are lower than those of SAW chains and depend on the wall
potential (Figure 6). However, since $\Phi=1/2$ applies to SAW as well as RW
chains, the $N_{S}/N^{1/2}$ vs $\tau N^{1/2}$ scaling with the proper choice
of $\tau$ is obtained in both cases for adsorption due to contact potentials
(Figure 7-8).

The adsorption behavior of homopolymers experiencing 10-4 potential and
contact potential is indistinguishable\cite{Metzger}. Our discussion suggests
that the adsorption of an annealed copolymer is analogous to the adsorption of
a homopolymer with $\epsilon_{eff}$ given by (\ref{I7}). Accordingly we expect
the adsorption of annealed homopolymers to be insensitive to the choice of the
potential and of the $z$ values used to define adsorbed monomers. Within the
accuracy of our simulation, this is indeed the case for the adsorption of
RW\ annealed copolymers where the data is collapsed by plotting $N_{S}%
/N^{1/2}$ vs $\tau N^{1/2}$ (Figure 9). It is however important to note that
for this case the $PH$ equilibrium is $z$ dependent. Accordingly $N_{SH}%
/N_{S}=K_{S}/(1+K_{S})$ and $N_{BH}/N_{B}=K_{B}/(1+K_{B})$ as specified by
(\ref{IV6}) and (\ref{IV8}) only capture the qualitative features of the PH
equilibrium but do not yield quantitative agreement because $\mu_{SH}^{0}$ and
$~\mu_{SP}^{0}$ as approximated via (\ref{IV4}) do not allow for the $z$
dependence of $U_{\mathrm{wall}}$. A quantitative agreement is achieved upon
dividing $\Delta\epsilon_{H}^{ad}$ or $\Delta\epsilon_{P}^{ad}$ by 1.5
allowing presumably for the average $U_{\mathrm{wall}}$ experienced within the
$z<2$ region (Figure 10).

\section{Discussion}

The adsorption behavior of annealed copolymers may conceivably differ with the
details of the model. Our discussion concerned the particular case of
non-cooperative two-state polymers involving unimolecular PH interconversion.
For this case, our results suggest that the annealed copolymers adsorb as
homopolymers once the appropriate effective adsorption energy per monomer at
the surface, $\epsilon_{eff}$, $\ $is identified. This $\epsilon_{eff}$
accounts for the average adsorption energy as well as the mixing entropy of
the surface PH monomers at equilibrium. The adsorption of two-state polymers
is described by the cross-over exponent $\Phi$ of homopolymer adsorption. In
this respect, their behavior is identical to that of quenched copolymers.
However, as opposed to quenched copolymers, where $\epsilon_{eff}$ as obtained
from the MW annealed approximation is determined by the fixed average $H/P$
fraction\cite{W, Mainz}, $\epsilon_{eff}$ in the annealed case depends on the
$PH$ equilibrium constants in the bulk and at the surface, $K_{B}$ and $K_{S}%
$. In other words $\epsilon_{eff}$ varies with the $H/P$ ratio which, in
contrast to the quenched case, depends on the temperature, the adsorption
energies and the bulk standard chemical potentials. This picture allows to
calculate the number of adsorbed monomers $N_{S}$, the $H/P$ ratio in the bulk
and at the surface as well as the overall $H/P$ ratio. The calculated values
are in good agreement with the simulation results concerning SAW and RW
experiencing contact potentials. Simulation of homopolymer chains suggest that
the form of the wall-monomer potential does not affect $\Phi$ and the scaling
behavior of $N_{S}.$ Since annealed copolymers adsorb as homopolymers
characterized by $\epsilon_{eff}$ one expects similar insensitivity to the
details of the potential. In contrast, the $H/P$ ratio is sensitive to the
range of the wall potential. Both features are apparent from our results
concerning the adsorption of RW annealed copolymers subject to a 10-4 wall
potential. As noted earlier, our analysis utilizes a simplified version of the
Karlstr{\"{o}}m model. In contrast to this model all monomer-monomer
interactions are identical. New features may well occur upon introducing
attraction between hydrophobic H monomers. In the case of free chain in the
bulk such interaction can lead to phase separation\cite{karl} and to polymer
collapse\cite{NY}. The preferential adsorption of H monomers together with HH
attraction may lead to adsorption induced collapse.

Our discussion thus far focused on the comparison between simulation results
of annealed and quenched copolymers. The comparison of the experimental
situation brings up a second issue. The experimental realization of quenched
copolymers is clear and studying the adsorption of the corresponding
homopolymers allows to deduce the adsorption energies of the different
monomers involved. The situation is more difficult in the case of annealed
copolymer as encountered in the modeling of NWSP. The corresponding one state
homopolymers do not exist. Furthermore, direct experimental evidence
concerning the molecular identity of the interconverting states is yet to
emerge. As a result the interaction parameters are obtained indirectly by
model dependent fitting of experimental data. With this in mind it is
important to recall the experimental evidence for two-state models. Two
observations are of special interest. One is their ability to predict the
qualitative features of the phase diagram and fit experimentally observed
phase boundaries\cite{gold, karl, tanaka, Veyt,Bekiranov1, kremer, Bekiranov2,
dorm}. The second concerns the stretching of PEO chains in atomic force
microscopy experiments. This reveals that different force laws characterize
the strong stretching regime in water and in hexadecane. In particular, the
chain extension in water exhibits a plateau characteristic of two-state
polymers\cite{Gaub}.

\begin{acknowledgement}
N.Y. benefited from fellowship No. 7662 from the Japan Society for the
Promotion of Science (JSPS) as well as from the hospitality of the theory
group of the ILL. Part of the numerical calculations in this work was carried
out on Altix3700 BX2 at YITP in Kyoto University.
\end{acknowledgement}

\pagebreak

\begin{figure}
\begin{center}
\includegraphics[width=0.6\textwidth]{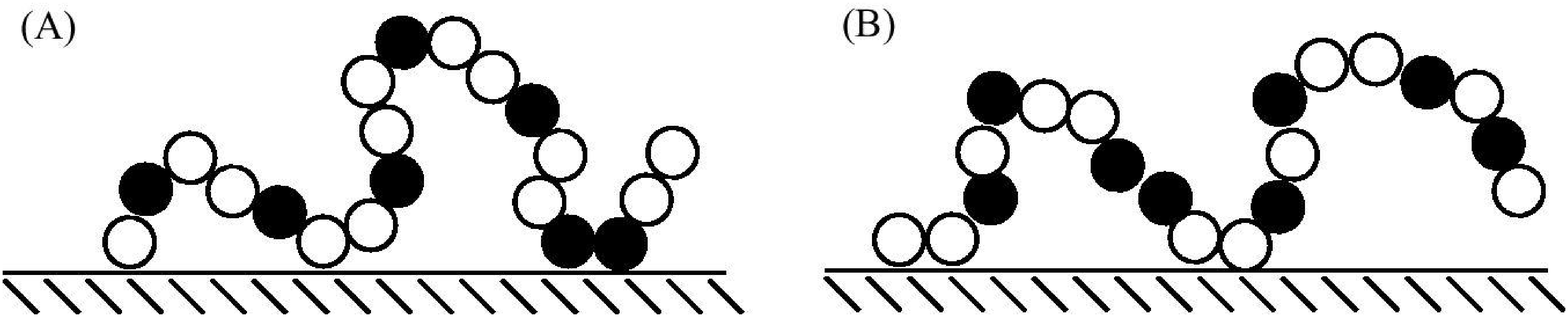}
\caption{
Two configuration of an adsorbed two-state polymer undergoing P(full
circles) H(empty circles) interconversion in the bulk and at the surface.
}
\end{center}
\end{figure}

\begin{figure}
\begin{center}
\includegraphics[width=0.6\textwidth]{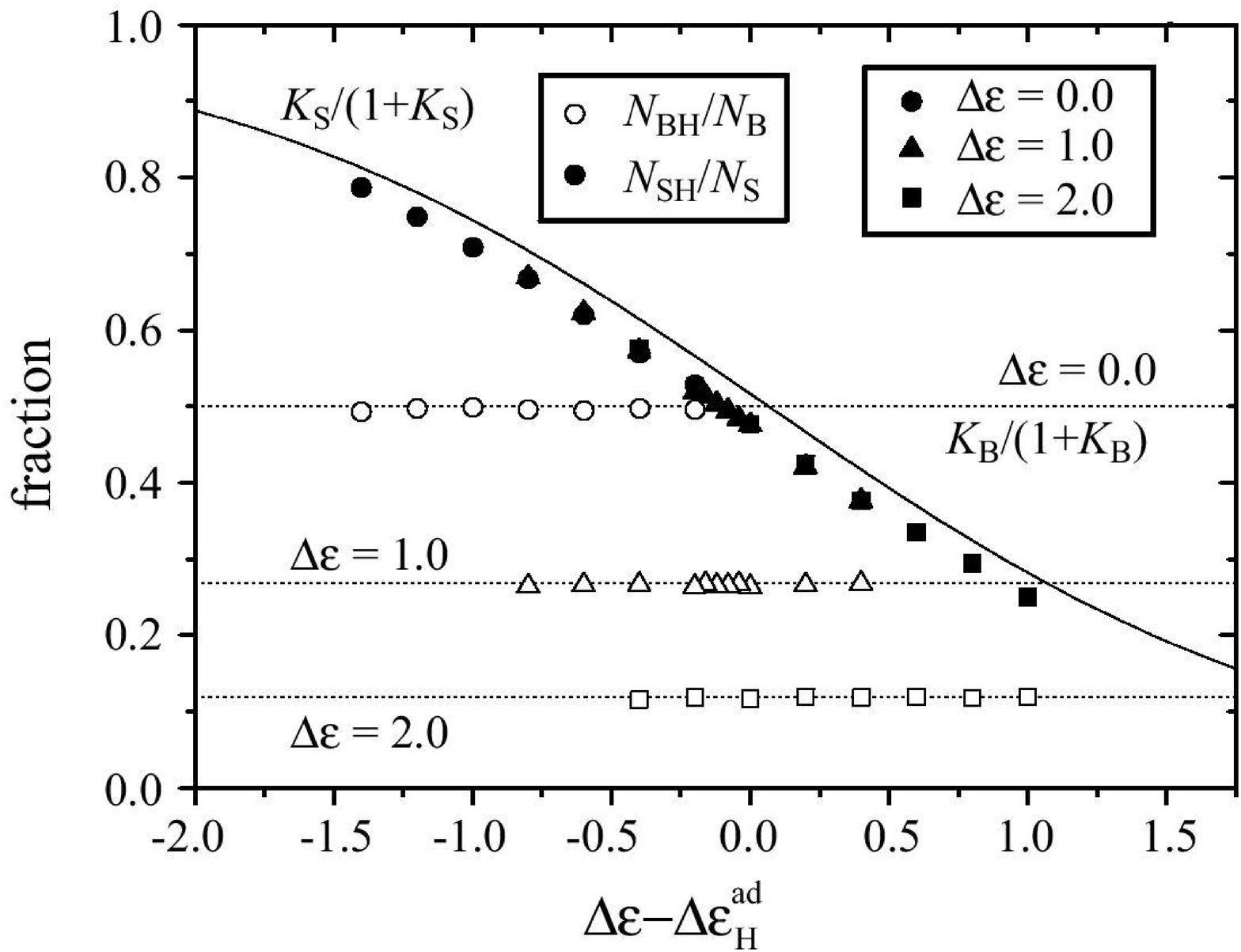}
\caption{
$N_{BH}/N_{B}$ and $N_{SH}/N_{S}$ as a function of a contact
potential $\Delta\epsilon-\Delta\epsilon_{H}^{ad}$ for various $\mu_{H}%
^{0}-\mu_{P}^{0}=kT\Delta\epsilon$ and the corresponding curves given by
(\ref{IV6}) and (\ref{IV8}).
}
\end{center}
\end{figure}

\begin{figure}
\begin{center}
\includegraphics[width=0.6\textwidth]{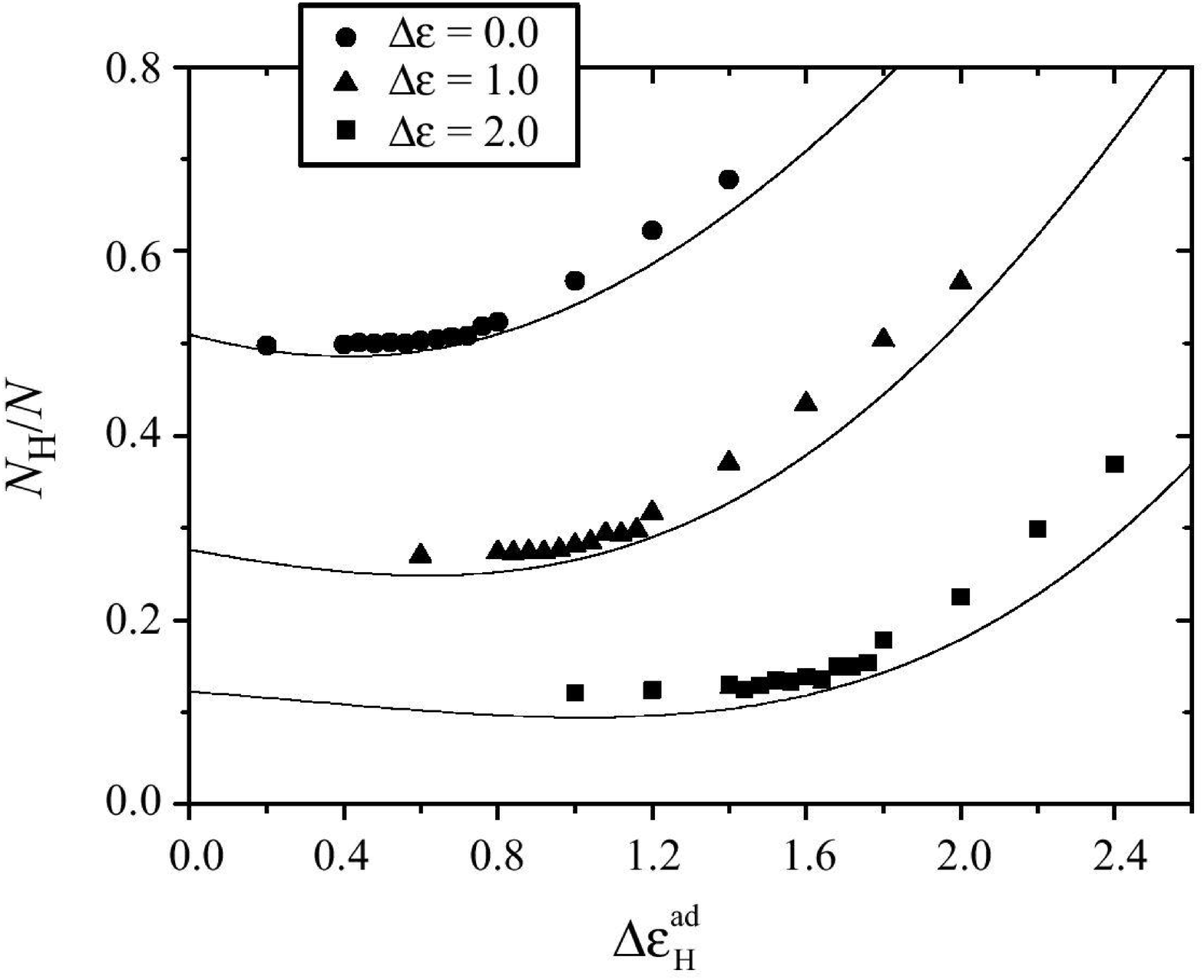}
\caption{
Plot of $N_{H}/N$ as a function of of a contact potential
$\Delta\epsilon_{H}^{ad}$ for various $\mu_{H}^{0}-\mu_{P}^{0}=kT\Delta
\epsilon$ and the corresponding curves given by (\ref{V2}).
}
\end{center}
\end{figure}

\begin{figure}
\begin{center}
\includegraphics[width=0.6\textwidth]{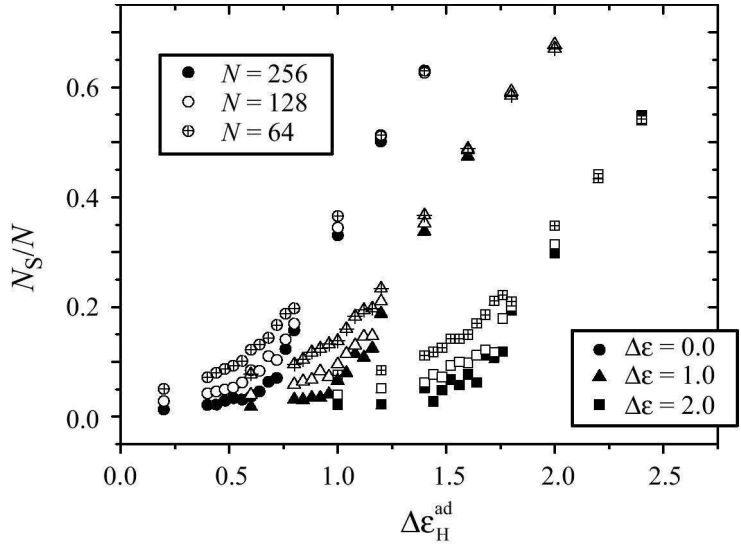}
\caption{
$N_{S}/N$ vs $\Delta\epsilon_{H}^{ad}$ of a contact potential for
self avoiding chains with various $\mu_{H}^{0}-\mu_{P}^{0}=kT\Delta\epsilon$
and $N$.
}
\end{center}
\end{figure}

\begin{figure}
\begin{center}
\includegraphics[width=0.6\textwidth]{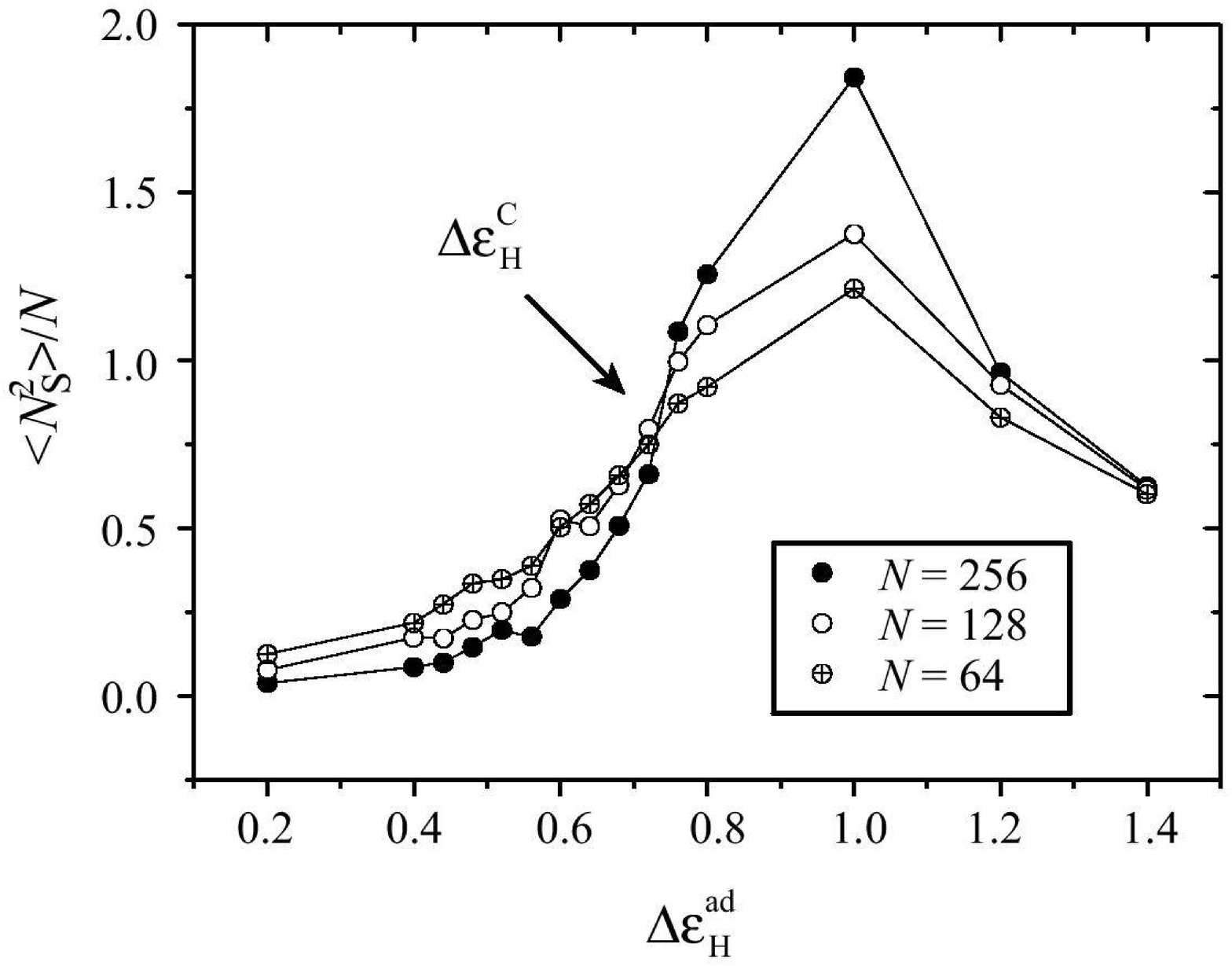}
\caption{
$N_{S}^{2}/N$ vs $\Delta\epsilon_{H}^{ad}$ for various $N$ and
$\Delta\epsilon=0$.
}
\end{center}
\end{figure}

\begin{figure}
\begin{center}
\includegraphics[width=0.6\textwidth]{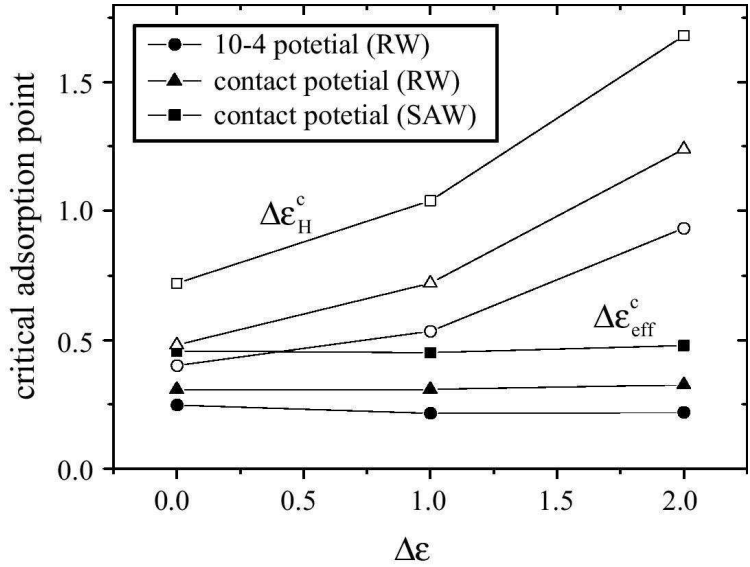}
\caption{
$\Delta\epsilon_{H}^{c}$ $\ $and the corresponding $\epsilon
_{eff}^{c}$ (as obtained by substituting $\Delta\epsilon_{H}^{c}$ into
(\ref{I7})) vs $\Delta\epsilon$.
}
\end{center}
\end{figure}

\begin{figure}
\begin{center}
\includegraphics[width=0.6\textwidth]{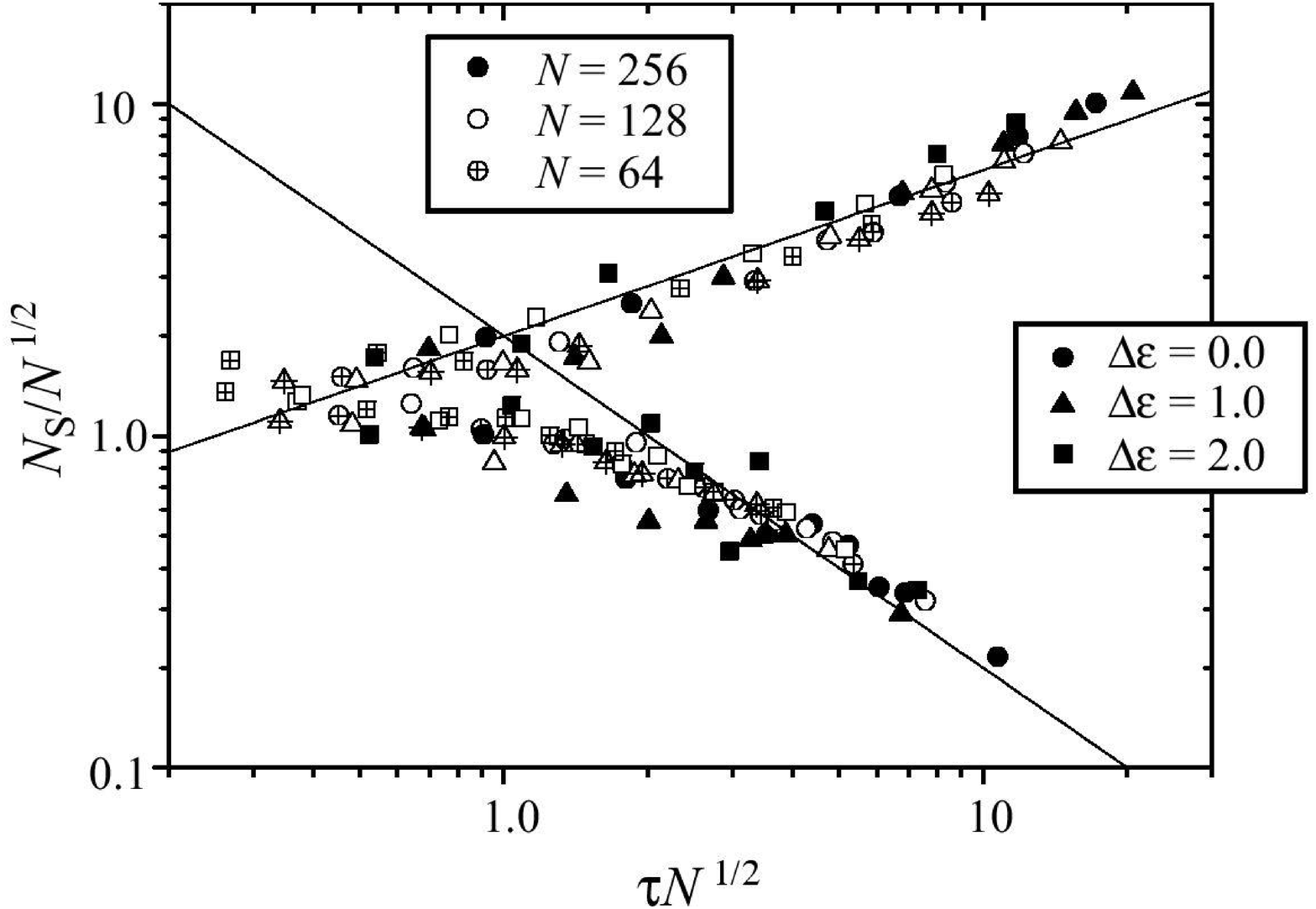}
\caption{
$N_{S}/N^{1/2}$ vs $\tau N^{1/2}$ for SAW experiencing a contact potential.
}
\end{center}
\end{figure}

\begin{figure}
\begin{center}
\includegraphics[width=0.6\textwidth]{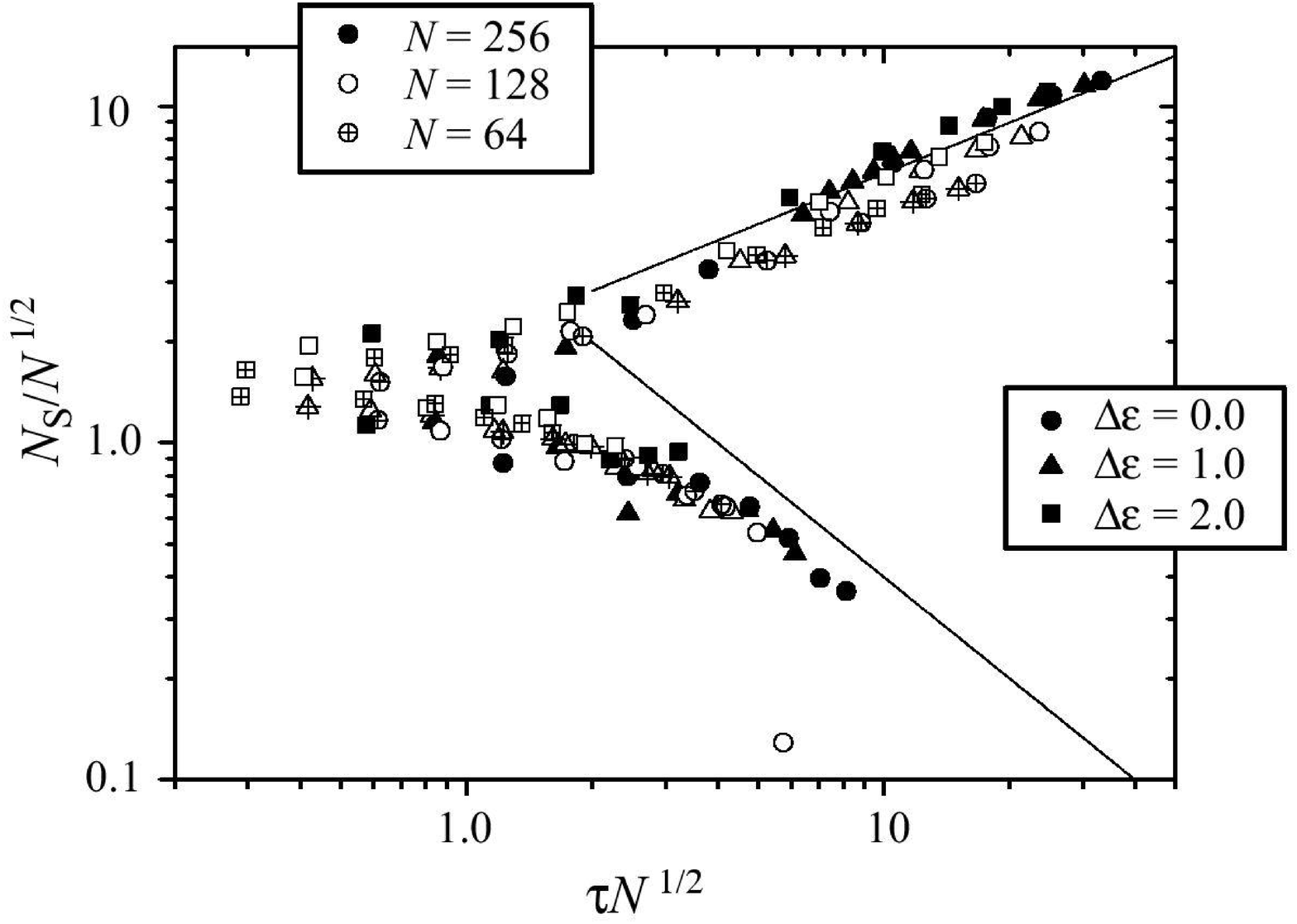}
\caption{
$N_{S}/N^{1/2}$ vs $\tau N^{1/2}$ for RW experiencing a contact potential.
}
\end{center}
\end{figure}

\begin{figure}
\begin{center}
\includegraphics[width=0.6\textwidth]{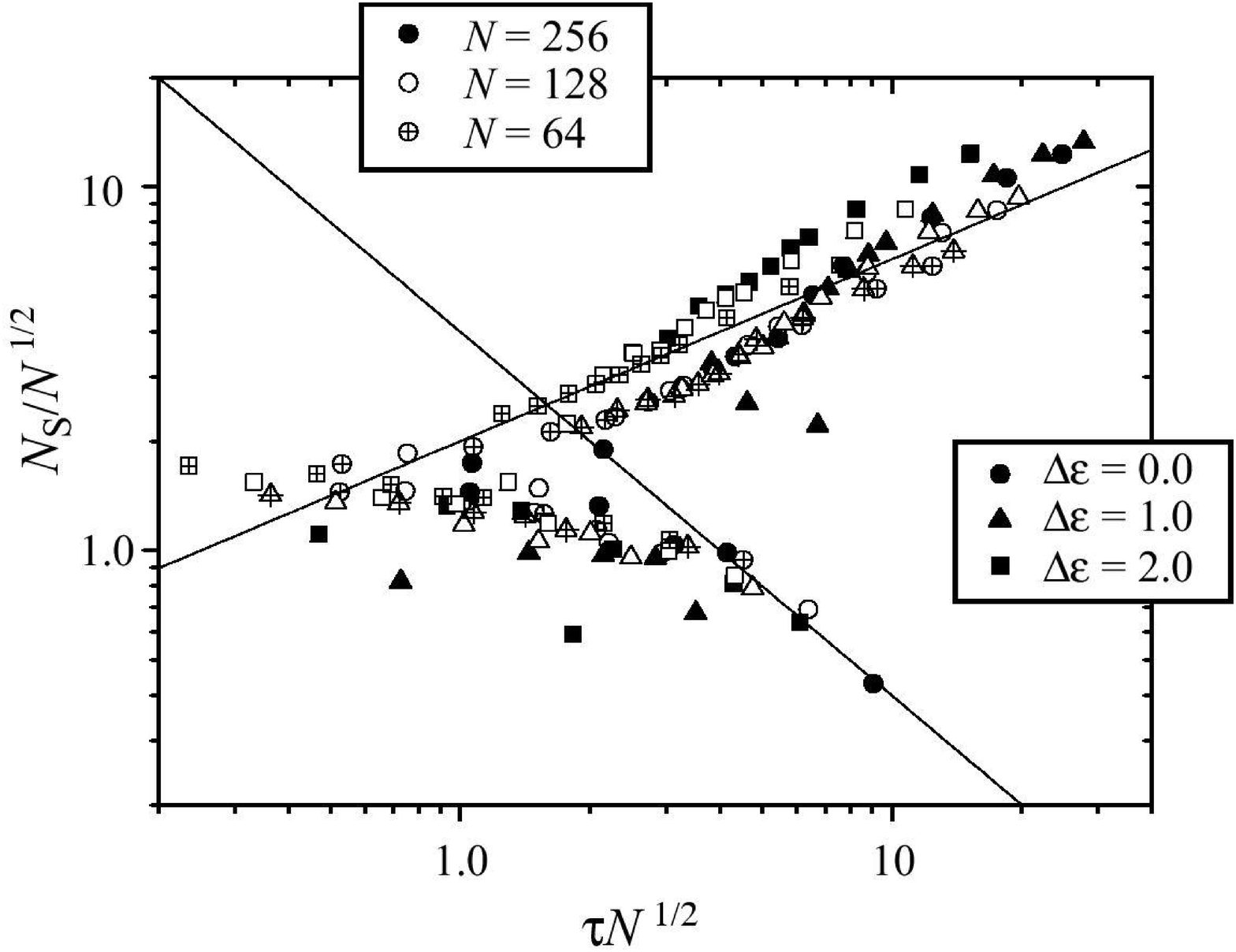}
\caption{
$N_{S}/N^{1/2}$ vs $\tau N^{1/2}$ for RW experiencing a 10-4 potential.
}
\end{center}
\end{figure}

\begin{figure}
\begin{center}
\includegraphics[width=0.6\textwidth]{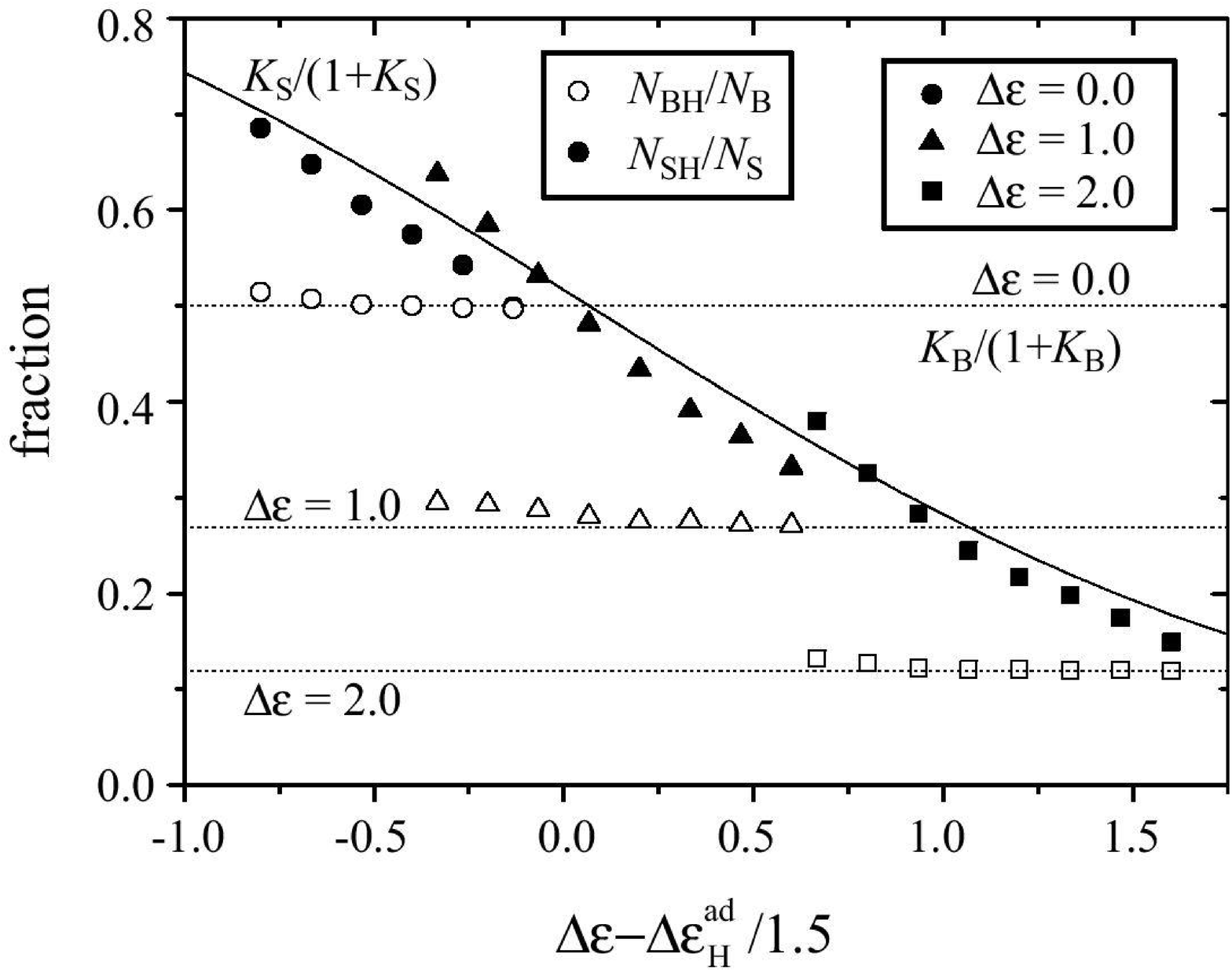}
\caption{
$N_{BH}/N_{B}$ and $N_{SH}/N_{S}$ as a function of a 10-4 potential
$\Delta\epsilon-\Delta\epsilon_{H}^{ad}/1.5$ for various $\mu_{H}^{0}-\mu
_{P}^{0}=kT\Delta\epsilon$ and the corresponding curves given by (\ref{IV6})
and (\ref{IV8}).
}
\end{center}
\end{figure}



\begin{thebibliography}{99}
\bibitem {foot}A fixed overall $H$ fraction is not gauranteed in the MW
annealed model especially for large $\epsilon$ \cite{W rev}.
\bibitem {foot2}Our discussion focused on the MW annealed approximation
because of its simplicity and proven utility in the analysis of simulation
data \cite{Z, Mainz}. We should add that the more general Morita approximation
\cite{W rev} is also obtainable from equation (\ref{II3}) upon choosing ${\mu
}_{P}^{0}/kT={\mu}_{SP}^{0}/kT=\lambda p-\ln(1-p)$, $-{\mu}_{H}^{0}%
/kT=\lambda(1-p)+\ln p$ and $-{\mu}_{SH}^{0}/kT=\epsilon_{MW}-\ln p$, ${\mu
}_{SH}^{0}/kT=\epsilon_{MW}+\lambda(1-p)+\ln p$.
\end{thebibliography}
\end{document}